\documentclass[twoside,11pt]{article}

\usepackage{blindtext}

\usepackage[abbrvbib, preprint]{jmlr2e}

\usepackage{lastpage}
\jmlrheading{26}{2025}{1-\pageref{LastPage}}{5/25; Revised 7/25}{9/25}{21-0000}{Deborah Pereg}

\usepackage{amssymb}
\usepackage{amsmath}
\usepackage{amsfonts}

\usepackage{float}
\usepackage{mwe}
\usepackage{graphbox}
\usepackage{setspace}
\usepackage{epstopdf}

\usepackage{xcolor}

\usepackage{bm}
\usepackage{acronym}

\usepackage[ruled,vlined]{algorithm2e}

\ShortHeadings{Embedding Capacity $\&$ Embedding Rate-Distortion}{Deborah Pereg et al.}

\firstpageno{1}

\begin{document}

\title{Information Theoretic Perspective on Representation Learning}

\author{\name Deborah Pereg \email deborah.pereg@supsi.ch \\
       \addr 
				Scuola Universitaria Professionale della Svizzera Italiana (SUPSI), \\
				Istituto Dalle Molle di Studi sull'Intelligenza Artificiale (IDSIA), \\
				Lugano, Switzerland.
				\AND
				Michael Wand 
				\email michael.wand@supsi.ch\\
				\addr
				Scuola Universitaria Professionale della Svizzera Italiana (SUPSI), \\
				Istituto Dalle Molle di Studi sull'Intelligenza Artificiale (IDSIA), \\
				Lugano, Switzerland.
			}

\maketitle

\begin{abstract}
An information-theoretic framework is introduced to analyze last-layer embedding, focusing on learned representations for regression tasks. We define representation-rate and derive limits on the reliability with which input-output information can be represented as is inherently determined by the input-source entropy. We further define representation capacity in a perturbed setting, and representation rate-distortion for a compressed output.
We derive the achievable capacity, the achievable representation-rate, and their converse. Finally, we combine the results in a unified setting. 
\end{abstract}

\section{Introduction}

Representation learning is one of the most investigated topics in machine learning in recent years. Previous works on representation learning focus mostly on classification downstream tasks, in which the representation ideally extracts the relevant information which is then ``compressed'' into the input classes \citep{Bengio:2013,Hinton:2020,Ben:2023,shwartz:2024,Luthra:2025}. 
Neural collapse \citep{Papyan:2020} is the observed phenomena that the representation of the training data (last layer training activations) collapses to a single representation for every class. 

Recent works \citep{Tishby:2015,shwartz:2017,shwartz:2018,kawaguchi:2023,shwartz:2024} introduced the information bottleneck (IB) principle in the context of supervised learning  and demonstrated that the convergence of DNNs' layers
follows the IB optimal bound. The IB principle assumes that an optimal representation would capture the relevant
features, and compress the input by dismissing the irrelevant parts which do not contribute to the prediction.
\cite{shwartz:2018} show that for a high dimensional input and typical input sequences, the mutual information up to a representation level (layer) $T$ controls the generalization gap. 

In previous work \citep{Pereg:2023A}, we derived theoretical bounds of sample complexity, based on the information-theoretic asymptotic equipartition property (AEP) \citep{ThomasCover:2006}. We showed that there exists a relatively small set that can empirically represent the input-output data distribution for learning and we formulated the relationship between the sample complexity, the empirical risk and the generalization error based on the typical-set properties. 

In this work, we focus on significant observations regarding the representation latent space in regression tasks.  
First, we observe that class-oriented collapsed representations are not necessarily advantageous for downstream regression tasks.
We then proceed to analyze the representation information-theoretic characteristics. We establish representation rate theorem for learned representations of bijective mappings (Theorem~\ref{Theorem 3}).
We further introduce representation capacity for learned representations under degraded or perturbed input environment as the maximal operational rate in bits per input (model-use) symbol at which information can be represented with arbitrarily low probability of error (Theorem~\ref{Theorem 4}). We continue to investigate representation rate-distortion in a compressed-output setting (Theorem~\ref{Theorem 5}). Lastly, we combine the results for a setting where both the input is noisy and the output is compressed (Theorem~\ref{Theorem 6}). 

\paragraph{Capacity in neural networks.}
We note that in the context of neural networks the term capacity is used for different meanings. 
To name a few, the term neuronal capacity \citep{Baldi:2018} is defined as the logarithm of the size (or volume) of the hypothesis class (i.e., number of functions a neural architecture can implement).
The capacity of feedforward neural networks \citep{Baldi:2019} defines cardinal capacity analogously. That is, the number of distinct functions (or with continuous weights, volume of function space) for feedforward nets of linear threshold gates.
Memory capacity \citep{Vershynin:2020} refers to the ability to memorize arbitrary labels as a function of architecture parameters (layers and activations). 

\paragraph{Remark 1.}
Note that despite obvious resemblance, there are major differences between Shannon's source coding and channel coding in communication theory \citep{Shannon:1948,ThomasCover:2006} and the theory established here for learning representations. In Shannon's source coding theory, we design an efficient sequence of symbols building the codebook, encoding information of the source, in an attempt to minimize the redundancy and compress the representation using fewer bits than in the original representation.  
The channel coding theorem establishes the rate at which we can transmit information over a channel with as little error as possible, inherently describing the maximum possible efficiency of error-correcting methods depending on the level of degradation imposed by the channel.  
Here, however, we have a given input-representation, which we could in principle modify, and a given structure for the embedding-representation (dimension and maximum number of bits per point-value) often determined by the designer of the neural network, and we are attempting to analyze the limits of the learned representation and its efficiency and reliability. The input to the model encoding the original sequence into its corresponding representation may be a degraded measurement contaminated by (physical) noise and/or digital noise (e.g., quantization noise, numerical errors). Alternatively the inputs may also be augmented views and/or transformations of a source signal, still corresponding with a shared embedding and the same output. 
As indicated above the representation does not necessarily compress the data, nor does it necessarily correct for ``channel" ambiguities caused by noise, varying measurements, et cetera. Thus, our goal is to reach a better understanding of the structure learned and to investigate the fundamental limitations imposed by the representation design, which in turn can shed light on the design of more efficient learning models.

\section{Preliminaries}

\subsection{Notations}
Let $X \in \mathcal{X}^{n\times 1}$, $Y \in \mathcal{Y}^{n\times 1}$, and $V \in \mathcal{V}^{d\times 1}$ be random variables
that obey stationary and ergodic probability distributions, and have a stationary coupling \citep{Gray:2009}
\footnote{A stationary random process is one for which the probabilities of an event are the same
regardless of when it occurs \citep{Gray:2009}.}.%
We consider a practical setting where $X$, $Y$ and $V$ are continuous
random variables represented in a ﬁnite precision machine where both $X$, $Y$ and $V$ are quantized into a ﬁnite number of
discrete values. 
We denote the joint probability mass function of $X$ and $Y$ as $P_{X,Y}(x,y)$, and their corresponding mutual information
is deﬁned as $I(X;Y) =\mathcal{D}(P_{X,Y}(x,y)||P_X(x)P_Y(y))\geq0$, where we have used the standard notation
$\mathcal{D}(p||q) \triangleq \sum P(u) \log \frac{P(u)}{Q(u)}$ for the Kullback-Leibler (KL) divergence between the probability mass functions $P$
and $Q$. Namely, in a discrete setting,
\begin{equation}\label{1.0}
I(X;Y)=\sum_{x,y} P_{X,Y}(x,y) \mathrm{log} \frac{P_{X,Y}(x,y)}{P_{X}(x)P_{Y}(y)}. 
\end{equation}

Equivalently, we assume a sample space that is a set $\Psi$ of paired objects $\Psi=\{\mathbf{x}_{i},\mathbf{v}_{i}\}_{i=1}^m$, where $\mathbf{x}_i \in  \mathcal{X}^{n\times 1}$ are sampled from $P_X(x)$ and paired with $\mathbf{v}_i \in  \mathcal{V}^{d\times 1}$ by a deterministic or stochastic mapping as ground truth. 
The notations $\mathbf{x}$ and $X$ are used below interchangeably. 

Hereafter, we use the notation $x^n$ to denote a sequence $x_1,x_2,...,x_n$, and $\mathbf{x}\in\mathcal{X}^{n \times 1}$ to denote a vector with $n$ entries. In information theory, a stationary stochastic process $u^n$ taking values in some finite alphabet $\mathcal{U}$ is called a source. More often than not, communication theory refers to discrete memoryless sources (DMS) \citep{Kramer:2008,ThomasCover:2006}. However, in many practical applications, signals, or local sections of them (such as image patches), can be modeled as ergodic sources belonging to some probability distribution forming statistical dependencies (e.g., a Markov random field (MRF) \citep{Weiss:2007}) describing the relations between data points in close spatial or temporal proximity.
To briefly summarize the AEP for ergodic sources with memory \citep{Austin:2017}, although the formal definition of ergodic process is somewhat complex, the general idea is simple: ``In an ergodic process, every sequence that is produced by the process is the same in statistical properties" \citep{Shannon:1948}. The symbol frequencies obtained from particular sequences generated by the process, approach a definite statistical limit, as the lengths of the sequences is increased.
More formally, we assume an ergodic source with memory that emits $n$ symbols from a discrete and finite alphabet $\mathcal{U}$, with probability $P_U(u_1,u_2,...,u_n)$.
We recall a theorem \citep{Breiman:1957}, here without proof.
\begin{theorem}[Entropy and Ergodic Theory \citep{Breiman:1957}] %
\label{Theorem 1.1} Let $(u_n)_{n \in \mathbb{Z}}$ be a stationary ergodic process ranging over a finite alphabet $\mathcal{U}$, then there is a constant $H$, defining the entropy rate of the source,
\begin{equation} 
H(U) = \lim_{n\rightarrow\infty} -\frac{1}{n} \log_2 P_U(u_1,...,u_n).
\end{equation} \end{theorem}

Intuitively, when we observe an ergodic source with memory over several entries, the uncertainty grows more slowly as $n$ grows, because once we know the previous source's entries, the dependencies reduce the overall conditional uncertainty. 
The entropy rate $H$, which represents the average uncertainty converges as the number of entries grows. This, of course, makes sense, as it is known that {\small $H(X,Y)\leq H(X)+H(Y)$}. In other words, the uncertainty of a joint event is less than or equal to the sum of the individual uncertainties.
The generalization of the AEP to arbitrary ergodic sources is as following \citep{Breiman:1953}.
\begin{theorem}[Shannon McMillan (AEP)\citep{Breiman:1953}]
\label{Theorem 1.2}
For $\epsilon>0$, the typical set $A^n_\epsilon$ with respect to the ergodic process $P_U(u)$ is the set of sequences $\mathbf{u}=(u_1,u_2,...,u_n)\in \mathcal{U}^n$ obeying
\begin{enumerate}
\item $\lim_{n \rightarrow \infty} \mathrm{Pr}[\mathbf{u} \in  A^n_\epsilon]=1$.
\item $2^{-n(H+\epsilon)} \leq P_U(\mathbf{u}) \leq 2^{-n(H-\epsilon)}$.
\item $(1-\epsilon)2^{n(H-\epsilon)} \leq |A^n_\epsilon| \leq 2^{n(H+\epsilon)}$, for $n$ sufficiently large.
\end{enumerate}
$|A|$ denotes the number of elements in the set $A$, and $\mathrm{Pr}[\mathcal{A}]$ denotes the probability of the event $\mathcal{A}$ .
\end{theorem}
In other words, if we draw a random sequence $(u_1,u_2,...,u_n)$, when $n$ is large enough, the typical set occurs approximately with probability 1. All elements of the typical set $A^n_\epsilon$ are approximately equally probable, and the number of elements of the typical set is approximately {\small $2^{nH}$}. This property is called the asymptotic equipartition property (AEP). In information theory the AEP is considered as the analog of the law of large numbers \citep{ThomasCover:2006}. %
The notion of a typical sequence was first introduced in 1948 by Shannon in his paper ``A Mathematical Theory of Communication'' \citep{Shannon:1948}.
Intuitively, the typical sequences $u^n$ are the sequences whose
\textit{empirical} probability mass function is close to $P_U(\cdot)$.

The above results lead to further information-theoretic interpretations and insights, in relation with channel coding and source coding. Notably, the capacity of a channel defines the optimal transmission rate of communication over a channel $P_{Y|X}(y|x)$. Shannon's channel coding theorem \citep{Shannon:1948} states that channel capacity is $C(P_{Y|X})=\max_{P_X}I(X;Y)$. 
In rate-distortion theory, we are trying to compress the source input under a constraint on the distortion. 
The optimal compression rate is $R(D)=\min_{P_{\hat{X}|X}}I(X;\hat{X})$ s.t. $\ Ed(\hat{x},x)\leq D$, where $x$ is the source, $\hat{x}$ is the decoded signal, $d(\hat{x},x)$ is a distortion measure and $D$ is a given distortion. 
Thus, in channel transmission, we wish to ﬁnd the largest set of codewords that have a large
minimum distance between codewords, whereas in rate-distortion, we try
to ﬁnd the smallest set of codewords that covers the entire space \citep{ThomasCover:2006}. 
So far, the direct connection with representation learning has been an open problem. In the general neural-network problem setting, given an input-output distribution, we are trying to design a system that captures the mapping between them, by learning from fewer examples as possible (lower sample complexity). We have previously shown that the generalization error bound depends on the sample complexity relative to a given input-output mutual information \citep{Pereg:2023A}. In this work we focus on characterization of information-theoretic properties of learned representations particularly in regression tasks, and establishing bounds determining possible implications in this context.

\subsection{Classification vs Regression}

Consider a mixture model $P_X(\mathbf{x})=\sum^M_{k=1}P_{\theta}(\theta_k)P_{X|\theta}(\mathbf{x}|\theta_k)$, such that $\mathbf{x}\in\mathcal{X}^{n \times 1}$, and $\theta$ represents one of $M$ underlying \textit{classes}. We further assume a forward \textit{regression} model $P_{V|X}(\mathbf{v}|\mathbf{x})$ independent of $\theta$. 
Accordingly, $P_V(\mathbf{v})=\sum^M_{k=1}P_{\theta}(\theta_k)P_{V|\theta}(\mathbf{v}|\theta_k)$.
Denote $\mathcal{F} : \mathcal{X}^n \rightarrow \mathcal{Z}^q$ as a function that maps input samples to $q$-dimensional embedding vectors $\mathbf{z}_i \in \mathcal{Z}^{q \times 1}$.
The embedding function $\mathcal{F}$ is trained via supervised learning or self-supervised contrastive learning, using a dataset $ \Psi = \{(\mathbf{x}_i,\mathbf{v}_i,c_i)\}^m_{i=1}$, where $c_i$ is the corresponding class label. 
In the self-supervised training the model only learns from the inputs $\{\mathbf{x}_i\}^m_{i=1}$, but not the corresponding class labels $c_i$. 

\begin{proposition}[Class-oriented representation-collapse for regression]
\label{Prop1}
Collapsed representations learned by employing self-supervised contrastive learning align with downstream classification tasks, but are inadequate for regression tasks.
\end{proposition}
That said, clustered representations that have not fully-collapsed to their mean, could still have sufficient variability and could potentially reliably represent the different output signals in every class.

\begin{proof}
\cite{Papyan:2020} showed that in classification deep neural networks, the top-layer feature embeddings of training samples of each class tend to cluster around respective class means, which are maximally distant from
each other. This phenomenon is generally considered desirable for classification tasks and segmentation \footnote{segmentation is considered here as classification of patches.} tasks because max-margin classifiers exhibit better generalization guarantees \citep{Ben:2023,Anthony:2009}, and clustered embedding spaces are useful for few-shot transfer learning \citep{Goldblum:2020, Galanti:2022}.

Previous work \citep{Luthra:2025} showed that when $\mathcal{F}$ is a global minimizer of the negatives-only supervised contrastive loss,
\begin{equation}
\mathcal{L}_{\mathrm{NSCL}}=-\frac{1}{K^2m} \sum^K_{l1,l2=1} \sum_{i=1}^m   \log  \Bigg( \frac{\mathrm{exp}(\mathrm{sim}(\mathbf{z}^{l1}_i,\mathbf{z}^{l_2}_i))}{\sum^K_{l_3=1} \sum_{i \neq j} \mathrm{exp}(\mathrm{sim}(\mathbf{z}^{l1}_i,\mathbf{z}^{l_3}_j))}\Bigg),
\end{equation}
where $\mathrm{sim}(\mathbf{z}_1,\mathbf{z}_2)=\frac{<\mathbf{z}_1,\mathbf{z}_2>}{\|\mathbf{z}_1\|_2\|\mathbf{z}_2\|_2}$ is the cosine similarity function, and $\mathbf{x}_i^l$ and $\mathbf{z}_i^l$ denote an input augmentation and its corresponding feature vector, 
the obtained representations obey:
\begin{enumerate}
\item Augmentation Collapse: For each class, and for every pair $l_1, l_2 \leq K$, we have 
$\mathbf{z}^{l_1}_{i}=\mathbf{z}^{l_2}_{i}$.
\item Within-Class Collapse: For any two samples $\mathbf{x}_i$ and $\mathbf{x}_j$ with the same label ($c_i = c_j$), their
representations coincide: $\mathbf{z}_i = \mathbf{z}_j$. Each class has a \textit{unique} class embedding.
\item Simplex Equiangular Tight Frame: Let $\{\mu_1, . . . , \mu_M\}$ denote the set of class-center embeddings.
Under the assumption that $\mathbf{z} \in \mathbb{R}^q$, these vectors form a simplex ETF in $\mathbb{R}^q$; specifically, they satisfy $\sum^M_{c=1}\mu_c=0$,
$\|\mu_c\|=\|\mu_{c'}\|$, $<\mu_c,\mu_{c'}>=-\frac{\|\mu_c\|_2^2}{M-1} \ \forall c \neq c'$.

\end{enumerate}

\cite{Luthra:2025} show that self-supervised decoupled contrastive loss \citep{Yeh:2022} follows similar behavior to the supervised model and thus exhibits neural collapse. 
Now, consider $\mathcal{F}$ is used for the regression task by adding a fully connected linear decoder and using the learned features.
That is $\hat{\mathbf{v}}= \mathbf{A}\mathcal{F}(\mathbf{x})$, where $\mathbf{A} \in \mathbb{R}^{d \times q}$. 
For a specific class, the regression outputs $\mathbf{v}_{i}^k \sim p(\mathbf{v}|\theta=\theta_k)$ within the same class $\theta_k$ vary:  
\begin{equation}
\mathbf{v}^k_i  \neq \mathbf{v}^k_j , \quad i \neq j, \ k=1,...,M.
\vspace{-10pt}
\end{equation} 
Denote the class unique embedding as $\mathbf{z}_{c_k}$. Since the learned features collapse, $\mathbf{z}_i|\theta_k=\mathbf{z}_j|\theta_k=\mathbf{z}_{c_k}$, all outputs within the same class collapse too, $\hat{\mathbf{v}_i}^k= \mathbf{A}\mathcal{F}(\mathbf{x}_i) =\mathbf{A}\mathbf{z}_{c_k}=\hat{\mathbf{v}_j}^k$.
This is true for any decoder $\mathcal{G}(\cdot)$ regardless of its architecture as all features in the same class are identical. 
\end{proof}

To a certain extent, this is a known property of classification neural nets, as the only information that must be preserved throughout the layers of the net is the class of the input signal.
Accordingly, the dimension of the representation is significantly smaller than the input dimensions, that is $q<<n$.
For example in ResNet-50 the input $\mathbf{x} \in \mathbb{R}^{224 \times 224 \times 3}$, thus $n \approx 2^{18}$, whereas $\mathbf{z} \in \mathbb{R}^{2048 \times 1}$, and $q=2^{11}$ \citep{He:2016}. In other words, it is assumed that the encoder compresses the input and maintains only the relevant information to distinguish between one class to another. 

In regressions tasks, such as image-to-image translation, architectures are normally over-parametrized.
Although in certain architectures the encoder contracts (such as Deep STORM \citep{Nehme:2018} and denoising autoencoder \citep{Rumelhart:1985}), in others the information from the input is passed directly to the decoder (e.g., UNET \citep{Ronneberger:2015}). In few-shot RNN \citep{Pereg:2020, Pereg:2023B, Pereg:2024B} the input $\mathbf{x} \in \mathbb{R}^{30 \times 1}$, thus $n \approx 2^5$, whereas $\mathbf{z} \in \mathbb{R}^{1000 \times 1}$, and $q \approx 2^{10}$.
It is indicated that $q>>n$ and that the representation is sparse. 
In those tasks the model must preserve all the information necessary to transform the input to the output space, and only discard noise and irrelevant information (in the sense that everything that is not an event - is noise). 
In early autoencoder designs \citep{Rumelhart:1985,Kramer:1991,Baldi:1989} the hidden layer is significantly smaller than the input size, (e.g., input of size $28 \times 28$ is mapped to 32 or 64 values) with the intention to compress the input and/or project it onto a lower-dimensional representation.  
Whereas modern denoising autoencoders \citep{Vincent:2008} are highly over-parametrized, e.g., an input of size  $28 \times 28$ is mapped to 1000-2000 sigmoid units. 
In machine learning literature, the embedding space is usually assumed to be continuous and finite, such that $\mathbf{z} \in \mathbb{R}^q , \ \|\mathbf{z}\| \leq 1$.
Here, we will take a different approach.  %
\paragraph{Confusion with compression and rate-distortion theory.}
The change of length of the block and the size of the alphabet, along with repeated abuse of terminology in the data-science literature, created a misleading misconception that the representation is always a compressed representation of the input. Whereas in most regression tasks each input must have a unique representation and a unique output, thus the ``codebook" size of the embedding must have sufficient ``capacity"\footnote{The term capacity is slightly abused here.} .  

\section{Representation Capacity}

\paragraph{Channel Capacity}
``What do we mean when we say that A communicates with B?
We mean that the physical acts of A have induced a desired physical state in B. 
This transfer of information is a physical process and therefore is subject to the
uncontrollable ambient noise and imperfections of the physical signaling
process itself. The communication is successful if the receiver B and the
transmitter A agree on what was sent." \citep{ThomasCover:2006}.
``A communication channel is a system in which the output depends
probabilistically on its input. It is characterized by a probability transition
matrix $p(y|x)$ that determines the conditional distribution of the output given the input"
 \citep{ThomasCover:2006}. 
In communication, channel capacity represents the maximum number of distinguishable signals
for $n$ uses of a communication channel.

\paragraph{Representation Capacity}
In our setting, that is, in representation learning, we assume a signal $\mathbf{x}\in\mathcal{X}^{n \times 1}$ that is an original ground truth signal (source) passing through a system, where the output of the system (channel) is the measurement $\mathbf{y}\in\mathcal{Y}^{n \times 1}$, which is the input to a neural net, to be represented by distinguishable $\mathbf{z}\in\mathcal{Z}^{q \times 1}$ in the embedding space (as illustrated in Fig.~\ref{fig2}).
Assuming that the embedding should represent distinguishable source signals $\mathbf{x}$, 
we ask: What is the maximum number of distinguishable signals per each input point value (pixel)?
For a system with source signal $X$ and output
$Y$, we can deﬁne the representation capacity $\mathcal{C}$ by
\begin{equation}
\mathcal{C} = \max_{P_X(\mathbf{x})} I(X; Y).
\end{equation}
Later we show that the representation capacity is the maximum rate at which
information can be reliably represented and recovered by the representation 
with a vanishingly low probability of error.

\begin{figure}
    \centering
    \includegraphics[width=0.87\textwidth]{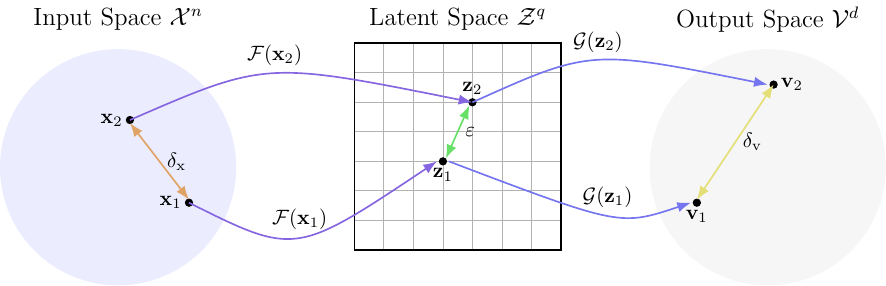}
		\caption{\footnotesize Illustration of the noise-free setting.}
		\label{fig1}
\end{figure}

Let us first consider a channel-free setting as illustrated in Fig.~\ref{fig1}.
Let $\mathbf{x} \in \mathcal{X}^{n \times 1}$ be an input signal, such that $|\mathcal{X}|$ is the cardinality of the input alphabet.
For example, in most image-related datasets the pixel values are originally integers $x_i \in [0,1,...,255]$ (normally normalized to the range $[-1,1]$ before further processing), thus $|\mathcal{X}|=2^8$.  
Let us assume a training set $\Psi=\{\mathbf{x}_{i},\mathbf{v}_{i}\}_{i=1}^m$, where $\mathbf{x}_i \in \mathcal{X}^{n\times 1}$ are sampled from $P_X(\mathbf{x})$ and paired with $\mathbf{v}_i \in \mathcal{V}^{d\times 1}$ by some function as ground truth. The learning system is trained to output a prediction rule $\mathcal{F}: \mathcal{X}^n \rightarrow \mathcal{V}^d$. Assume an algorithm 
that trains the predictor by minimizing the training error (empirical error or empirical risk).

\paragraph{Bijective transformation} 
For the sake of the following theoretical analysis we restrict the mapping $\mathcal{F}: \mathcal{X}^n \rightarrow \mathcal{V}^d$ to be a bijective (invertible) function. Namely, for every $v^d$, there is a single $x^n$ such that $v^d=g(x^n)$, and for every $x^n$, there is a single $v^d$ such that $x^n = g^{-1}(v^d)$.
In other words, every element $v^d$ is the image of one element of $x^n$. It is required that $x^n$ is \textit{unique}.
As we established in the Proposition \ref{Prop1}, for every two distinguishable outputs $\mathbf{v}_i\neq \mathbf{v}_j$, the corresponding embedding must be distinguishable too, $\mathbf{z}_i\neq\mathbf{z}_j$. Since every input must have an output, by assumption $|\mathcal{X}|^n \leq |\mathcal{V}|^d$. 

\paragraph{The Embedding Space}
More often than not, the embedding space is assumed to be a floating point in a continuous space such that
$\mathbf{z} \in \mathbb{R}^q , 0 \leq |\mathbf{z}| \leq 1$. Here, we will assume a finite digit representation such that $\mathbf{z}\in\mathcal{Z}^q$, where $|\mathcal{Z}| << \infty$. 

Considering the first classification network for MNIST LeNet-5 \citep{Lecun:2002}, the input images of size $32 \times 32$, when $x_i\in [0,1,...255]$.
Therefore, $|\mathcal{X}|^n= 256^{1024}=2^{8192}$. For the sake of simplification we will assume $\|z_i\| \leq 1$ represented by float32, and a rough estimate of $|\mathcal{Z}| \approx 2^{31}$ quantized numbers in this range. (If one is using TensorFloat-32, $|\mathcal{Z}| \approx 2^{10}$).
\textit{It is often assumed that the representation (feature) space reduces the dimensionality of the input because usually 
$n >> q$}.
Although $|\mathcal{Z}|>>|\mathcal{X}|$, in many cases the embedding space does not have the ``capacity'' to represent every point in the input ambient space, including those which are not part of the support of the input manifold (e.g., input manifold of digits, natural images, etc.). 
For example, if $q=128$, then we can only have at most $|\mathcal{Z}|^q= 2^{31*128}=2^{3968}$ representations. For the embedding space to have as many distinguable points as potentially in the full input ambient space $|\mathcal{X}|^n=2^{8192}$ we would require that $|\mathcal{Z}|=2^{64}$, which is unrealistic in practice. 
Accordingly, the Data Manifold Hypothesis (Olah, 2014) postulates that real-world datasets lie on a manifold
whose intrinsic dimensionality is considerably lower than the ambient input space. While numerous models
can completely fit training data, in order for the model to generalize well, it likely learns meaningful
representations of this underlying manifold.

That said, we note that saying that the embedding space is \textit{compressing} the input is generally a misleading statement.
For example, if an input image patch of size $64 \times 64$ where pixels are integers in the range $[0,255]$, is represented in the embedding space in a vector $\mathbf{z} \in \mathcal{Z}^{1024 \times 1}$ represented as a floating point with 32 bits. Then each input pixel is represented by 8 bits in the embedding space, which is the same number of bits per symbol as in the input representation space.
If the input image patch is smaller, say of size $16 \times 16$ where pixels are integers in the range $[0,255]$, represented in the embedding space in a vector $\mathbf{z} \in \mathcal{Z}^{1024 \times 1}$ represented in as a floating point with 32 bits, then each input pixel is represented by 128 bits in the embedding space which is a significantly larger number of bits.

\paragraph{The Embedding Separation Distance}
When the model is a bijective function every input has a corresponding distinguisbale representation point in the $\mathcal{Z}$-domain. 
Every input has an embedding that is \textit{separable} from other embeddings by a $q$-dimensional sphere. 
We will refer to the radius of that sphere as the \textit{embedding separation distance} denoted by $\varepsilon$.

\paragraph{Embedding Continuity}
Let us assume the mapping $\mathcal{F} : \mathcal{X}^n \rightarrow \mathcal{Z}^q$ is Lipschitz continuous\footnote{Only required for Theorem \ref{Theorem 7}.} with a constant $K_c>0$
\begin{equation}
\frac{\|\mathbf{z}_i-\mathbf{z}_j\|}{\|\mathbf{x}_i-\mathbf{x}_j\|}\leq K_c, \quad \forall \mathbf{z} \in \mathcal{Z}^q, \ \mathbf{x} \in \mathcal{X}^n,
\end{equation} 
where $\|\cdot\|$ denotes a norm. Thus requiring that the target function is locally smooth,
i.e., if we perturb the input $\mathbf{x}$ slightly
the embedding does not change much. 

\section{Theoretical Bounds}

We first assume a supervised learning algorithm with a training set $\Psi$,
where the input $x^n$ is sampled from an unknown distribution $P_X$ and paired with $v^d$ by some target function $g(\cdot)$,
to learn a predictor $h_{\Psi} : \mathcal{X}^n \rightarrow \mathcal{V}^d $.
The algorithm is designed to yield a model $h_{\Psi}$ that would ideally minimize the error over an unknown 
$P_{X,V}(\cdot)$ over $\mathcal{X}^n \times \mathcal{V}^d $, as the true error is not
available to the learner. The subscript $\Psi$ emphasizes that the output predictor depends on $\Psi$.
The learning process minimizes the empirical risk
\begin{equation}\label{2.1}
\mathcal{L}_{\Psi}(h_{\Psi}) = \frac{1}{m} \sum_{i=1}^m \ell(h_{\Psi}(\mathbf{x}_i),\mathbf{v}_i),
\end{equation}
where $0 \leq \ell(\hat{\mathbf{v}},\mathbf{v}) \leq 1$ is some loss function. 
The empirical error over the training set at the end of the training, for the specific trained predictor $h_{\Psi}$ is $\mathcal{L}_{\Psi}(h_{\Psi})\leq \Delta_m<<1$. The true error, or the \textit{generalization error}, in this setting is
\begin{equation}\label{2.2}
\mathcal{L}(h_{\Psi}) = E_{(x,v) \sim P_{X,V}}\big[\ell \big(h_{\Psi}(\mathbf{x}),\mathbf{v}\big)\big],
\end{equation}
where $E_{(x,v) \sim P_{X,V}}[\cdot]$ denotes the expectation over $P_{X,V}$.
The ability of a trained learner to generalize well is guaranteed by the upper bound on the generalization error,
$\mathcal{L}(h_{\Psi})\leq \Delta$ \citep{Pereg:2023A}. 

\begin{definition}[Embedding Code]
A source code $C_{\mathrm{z}}(x)$ for a random source sequence $x^n$ is a mapping $C_{\mathrm{z}}(x): \mathcal{X}^n \rightarrow \mathcal{Z}^q$. 
\end{definition}

\begin{definition}[Embedding representation rate]
A source code $C_{\mathrm{z}}(x)$ for a random source sequence $x^n$ 
and a rate of $\mathcal{R}$ bits per source symbol
is a mapping $C_{\mathrm{z}}(x): \mathcal{X}^n \rightarrow \mathcal{Z}^q$.
The embedding produces $\mathcal{R}$ bits per input symbol. 
Namely, $2^{n\mathcal{R}}=|\mathcal{Z}|^q$.
Denote, $\mathcal{Q}_{\mathrm{z}}= \log_2 |\mathcal{Z}|^q$.
Therefore, 
\begin{equation}
\mathcal{R} =\frac{q \log_2 |\mathcal{Z}|}{n} = \frac{\mathcal{Q}_{\mathrm{z}}}{n} . 
\end{equation}
\end{definition}
For example, in the above MNIST LeNet-5 \citep{Lecun:2002}, $\mathcal{R}=\frac{128\log_2(2^{31})}{32*32}=3.875$ bits per input pixel.
As stated above, the input pixels were originally represented by 8 bits per pixel. Therefore, the embedding does not have the ``capacity"\footnote{the use of the term capacity here, as in most machine learning literature, is slightly misleading.} to represent every possible input in the full support (input ambient space) with a distinguishable representation, however it may cover the empirical support defined by the typical set, as described in Theorem \ref{Theorem 3} below.  

\subsection{Representation rate in the noise-free setting}

\begin{theorem}[Embedding representation rate for a bijective mapping]
\label{Theorem 3}
Let $g(\cdot)$ be a deterministic bijective function. %
Let $\Psi = \big\{ \{\mathbf{x}_{i},\mathbf{v}_{i}\}_{i=1}^m : 
\mathbf{x}_i \sim P_X, \quad \mathbf{v}_i= g(\mathbf{x}_i), \ \mathbf{x}_i \in \mathcal{X}^{n\times 1},\ \mathbf{v}_i \in \mathcal{V}^{d\times 1} \big\}$ be a training set that is generated by randomly drawing samples from $P_X$ and labeling them by the target function $g(\cdot)$. Let $h_{\Psi} : \mathcal{X}^n \rightarrow \mathcal{V}^d$ be a trained predictor such that, $h_{\Psi}=\mathcal{F}(\mathcal{G}(\mathbf{x}))$, where $\mathcal{F}: \mathcal{X}^n \rightarrow \mathcal{Z}^q$, and $\mathcal{G} : \mathcal{Z}^q \rightarrow \mathcal{V}^d$. 
For $n$ sufficiently large, if the embedding space obeys
\begin{equation}
\mathcal{Q}_{\mathrm{z}} \geq n H(X), 
\end{equation}
in other words,
\begin{equation}
\mathcal{R} \geq H(X), 
\end{equation}
then there exists a trained predictor $\hat{\mathbf{v}}=h_{\Psi}(\mathbf{x})=\mathcal{G}(\mathcal{F}(\mathbf{x})), \ \mathbf{z}=\mathcal{F}(\mathbf{x})$,  such that the probability of error $P_e \triangleq Pr[h_{\Psi} (\mathbf{x}) \neq \mathbf{v}] \rightarrow 0, \ \forall \mathbf{x} \sim P_X$. That is, the generalization error $\mathcal{L}(h_{\Psi}) \rightarrow  0$. 

Conversely, if $\mathcal{R} < H(X) - \epsilon$ then the probability of error \\ $P_e \triangleq Pr[h_{\Psi} (\mathbf{x}) \neq \mathbf{v}] \rightarrow 1, \ \forall \mathbf{x} \sim P_X$.
\end{theorem}
\textit{Proof.} See Appendix \ref{AppA}.

\paragraph{Remark 2.} The proof deduces that for every $\mathbf{x} \in A^n_\epsilon(P_X) $ there is a corresponding unique output $\mathbf{v}$, hence
\begin{equation}\label{5.1.11}
d \log_2|\mathcal{V}| \geq nH(X).
\end{equation} 

Note that assuming that all inputs are equiprobable (either on the basis of the AEP, or by the nature of the input distribution), and thus all representations are equiprobable, $H(Z)=\mathcal{Q}_{\mathrm{z}}$, which is the maximum entropy of the representation achieved when all possible representations are uniformly distributed.

\paragraph{Remark 3.} Note that these theorems are independent of the hypothesis class of $h_{\Psi} (\mathbf{x})$.
That is, if the rate is insufficient at the representation level, no architecture can compensate for that. 

\subsection{Representation rate in a noisy environment}

\begin{figure}
    \centering
    \includegraphics[width=0.97\textwidth]{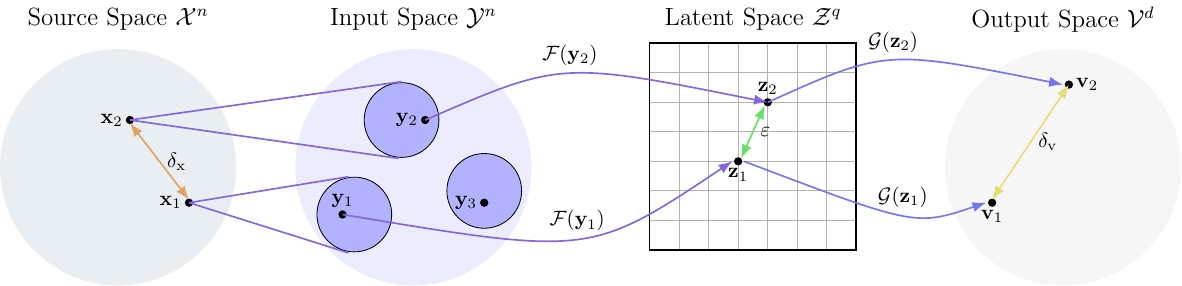}
		\caption{\footnotesize Illustration of a noisy environment problem setting.}
		\label{fig2}
\end{figure}

Now, let us assume a discrete source $P_X(\cdot)$ that emits i.i.d sequences $x^n$ of symbols (for example: patches in an image, frames of an audio signal, etc.), the estimator has access only to the observed signal $y^n$, which is a transformed form of $x^n$ according to the conditional distribution $P_{Y|X}(\mathbf{y}|\mathbf{x})$. 
For example, $y^n$ could be a degraded signal originating in $x^n$, as a result of additive noise, linear or nonlinear degradation and/or quantization, where the relationship between $x^n$ and $y^n$ could be linear or non-linear, with or without additive noise, such that generally $y^n=u(x^n)+e(x^n)$, where $u(\cdot)$ and $e(\cdot)$ are functions of $x^n$. 
Indeed, most real-life practical signal-to-signal models assume an input measurement $\mathbf{y}$ that could be noisy and/or degraded, yet the output should correspond with the central (and/or average) clean signal, denoted here as $\mathbf{x}$ (e.g., EMG to speech \citep{Hou:2024}, medical imaging reconstruction \citep{Young:2024}, speech to text \citep{Baevski:2020}, and so on) (see Figure~\ref{fig2}). 
Importantly, $P_{Y|X}(\cdot|\mathbf{x})$ corresponds with any heuristic stage generating a  noisy or distorted or transformed measurement from the original ground truth $\mathbf{x}$, when $\mathbf{y}$ are the inputs to the neural net.
This stage can also correspond with any preprocessing stage applied to the signal, which yields the input to the neural net, such as: rescaling, normalization, clipping and quantization, as well as any augmentations of the input, such as different views from different angles, different contrast, colors, etc., which are to be mapped to the same output $\mathbf{v}$, thus correspond with the same source $\mathbf{x}$. %
For the sake of the following theoretical analysis we restrict the learned mapping $h_{\Psi}: \mathcal{Y}^n \rightarrow \mathcal{V}^d$ to be a surjective function. Namely, for every $v^d$, there is a $y^n$ such that $h(y^n) = v^d$. In other words, every element $v^d$ is the image of at least one element of $y^n$. It is not required that $y^n$ be unique. The relation $\mathbf{x} \rightarrow \mathbf{v}$ is a bijection, and the input to the estimator is $\mathbf{y}\sim P_{Y|X}(\cdot|\mathbf{x})$.
In the presence of noise this condition can only be met in practice if the noise's power is under a certain threshold. 
In an inverse problem where the predictor is trained for signal restoration
$v^d=x^d$ and given $y^d$, we produce an estimate $\hat{x}^d$ (n=d), 
however, the following proofs are not restricted to this framework. 
Further assume a training set $\Psi=\{\mathbf{y}_{i},\mathbf{v}_{i}\}_{i=1}^m$, where $\mathbf{y}_i \in \mathcal{Y}^{n\times 1}$ are sampled from $P_{Y|X}$ and paired with $\mathbf{v}_i \in \mathcal{V}^{d\times 1}$ by some function as ground truth. The learning system is trained to output a prediction rule $h_{\Psi}: \mathcal{Y}^n \rightarrow \mathcal{V}^d$. Assume an algorithm 
that trains the predictor by minimizing the training error (empirical error or empirical risk). 

\begin{definition}[Effective Support of the Embedding Space]
The non-zero support of the embedding of a specific model $\mathbf{z}=\mathcal{F}_{\Psi}(\cdot)$ is defined as 
\begin{equation}
\mathcal{S}(\mathbf{z})= \Big\{ i\in \big\{1,2,...,|\mathcal{Z}|^q \big\} \ : \  \mathbf{z}_i \in \mathcal{Z}^q, \ \mathbf{z}_i\neq \mathbf{0} \Big\}.
\end{equation}
We denote the cardinality of the non-zero support,
\begin{equation}
\tilde{\mathcal{Q}}_{\mathrm{z}} \triangleq \log_2 \big|\mathcal{S}(\mathbf{z})\big|. 
\end{equation}
\end{definition}

Let $\mathcal{B}_{\mathrm{x}} = \{\mathbf{x}_i, \ i\in \mathcal{M}\}$ be a set of distinct $M=|\mathcal{M}|$ sequences in $\mathcal{X}^{n\times 1}$ that are resolved from the inputs $\mathbf{y}$ and mapped to their corresponding $\mathbf{v}_i$, and $M \triangleq 2^{n\mathcal{R}}$.

\begin{theorem}[Embedding Capacity]
\label{Theorem 4}
Let $\Psi$ be a training set that is generated by randomly drawing samples from $P_{Y|X}$ and labeling them by the target function such that,  
$\Psi = \big\{ \{\mathbf{y}_{i},\mathbf{v}_{i}\}_{i=1}^m : 
\mathbf{y}_i \sim P_{Y|X}(\cdot|\mathbf{x}), \ \ \mathbf{x}_i \in \mathcal{X}^{n\times 1},\ \mathbf{y}_i \in \mathcal{Y}^{n\times 1},\ \mathbf{v}_i \in \mathcal{V}^{d\times 1}\big\}$. 
Let $h_{\Psi} : \mathcal{Y}^n \rightarrow \mathcal{V}^d$ be a trained predictor such that, $h_{\Psi}=\mathcal{F}(\mathcal{G}(\mathbf{y}))$, where $\mathcal{F}: \mathcal{Y}^n \rightarrow \mathcal{Z}^q$, and $\mathcal{G} : \mathcal{Z}^q \rightarrow \mathcal{V}^d$. 
For $n$ sufficiently large, if 
\begin{equation}
\mathcal{R} < I(X;Y).
\end{equation}
then there exists a trained predictor $h_{\Psi}$ such that the probability of error \\ $P_e \triangleq Pr[h_{\Psi} (\mathbf{y}) \neq \mathbf{v}] \rightarrow 0, \ \forall \mathbf{y} \sim P_{Y|X}(\cdot|\mathbf{x})$.\\ 
When $\mathcal{G}$ is an injective mapping, the effective support of the embedding space obeys
\begin{equation}
\tilde{\mathcal{Q}}_{\mathrm{z}} < n I(X;Y). 
\end{equation}%
Conversely, if $P_e \triangleq Pr[h_{\Psi} (\mathbf{y}) \neq \mathbf{v}] \rightarrow 0, \ \forall \mathbf{y} \sim P_{Y|X}(\cdot|\mathbf{x})$, then $\mathcal{R} < I(X;Y)$.
\end{theorem}
\textit{Proof.} See Appendix \ref{AppB}.

\paragraph{Remark 4.} We note that when $\mathcal{G}$ is injective, for a perfect representation we have $H(X)=H(Z)=H(V)$ since the representation and the output are a bijection of the source, as well as $I(X;Y)=I(Z;Y)=I(V;Y)$. %

\paragraph{Remark 5.} The term ``effective support of the embedding space'' refers to the non-zero support,
indicating that the upper bound of $\tilde{\mathcal{Q}}_\mathrm{z}$ does not limit the practical utility of an embedding space of higher representation rate, but it defines its inherent limited ability to capture a limited number of representations.
A representation of higher dimension, or of larger alphabet, would be able, in principle, to reliably represent more signals. Yet, the number of distinguishable signals in this setting is limited by $\mathcal{R} < I(X;Y)$.  
However, if $\mathcal{G}(\mathbf{z})$ is an injective mapping and the effective support $\tilde{\mathcal{Q}}_{\mathrm{z}} > n I(X;Y)$, it implies that different $\mathbf{y}$'s originating in the same $\mathbf{x}$ are mapped to different $\mathbf{v}$'s, and that the mapping is erroneous. 

\paragraph{Remark 6.} The assumption that $\mathcal{G} : \mathcal{Z}^q \rightarrow \mathcal{V}^d$ is an injective mapping may seem too restrictive, but recent work indicates that in regression tasks $\mathcal{G}$ is invertible on the collapsed latent subspace \citep{Ross:2024}. 
It is shown that the last-layer feature vectors collapse to the subspace spanned
by the $d$ principal components of the feature vectors, where $d$ is the dimension of
the targets, and also to the subspace spanned by the last-layer weight vectors, comprising $\mathcal{G}$ in our setting.

\paragraph{Remark 7.} Despite the aforementioned significant differences between transmission rate in communication theory and the established theory here, and although the embedding rate does not directly capture an efficiency of error-correcting methods, it does bound the limits to which the learned model can correct for errors or the imperfections determined by the transformation $P_{Y|X}(\cdot|\mathbf{x})$. It is indicative of the extent to which the learned mapping can exploit structural redundancy to reconstruct the correct output.  

\begin{definition}[Representation Capacity]
The representation capacity is
\begin{equation}
\mathcal{C} = \max_{P_X(\mathbf{x})} \tilde{\mathcal{Q}}_{\mathrm{z}} = \max_{P_X(\mathbf{x})} \ I(X;Y),
\end{equation}
where the maximum is taken over all possible input distributions $P_X(\mathbf{x})$.
\end{definition}
Thus, the operational deﬁnition of representation capacity is the
highest rate in bits per model-use at which information can be represented with
arbitrarily low probability of error.
\textit{The representation capacity inherently depends on how many input signals are to be mapped to the same embedding point.}

Example 1. Consider $\mathbf{y} = \mathbf{x} + \mathbf{e}$,
where $\mathcal{X}^n = \{\mathbf{x}_1, \mathbf{x}_2, ..., \mathbf{x}_{M} \}$
and $\mathcal{E}^n = \{\mathbf{e}_1, \mathbf{e}_2, ..., \mathbf{e}_{P} \}$ are independent and both uniformly distributed. Further assume that the map $\mathbf{y} \rightarrow \mathbf{x}$ is surjective.
Since in this case $Y^n$ is uniform on its support with $P_{Y^n}(y_i^n)=\frac{1}{M P}, \ i=1,...,MP$, we have 
\begin{equation*}
I(X;Y) = H(Y)-H(Y|X) = H(Y)- H(E) = \frac{1}{n} (\log_2 MP - \log_2 P)= \frac{1}{n} \log_2 M = H(X). 
\end{equation*}

Example 2. Additive Gaussian noise with discrete input probability mass function.
For the sake of simplicity, in this example we will consider $\mathbf{Y} \in \mathbb{R}^n$. 
We consider a discrete random signal $X^n$ taking values in a finite alphabet $\Omega \ \subseteq \mathcal{X}^n$. The number of symbols in the constellation is $|\Omega|=M$, with probability $P_{X^n}(x_i^n)=\frac{1}{M}, \ i=1,...,M$ \citep{Delsad:2023}.
The degraded input signal $Y^n$, where
$\mathbf{y} = \mathrm{snr} \ \mathbf{x} + \mathbf{e}$ and $\mathbf{e}\sim \mathcal{N}(0,\mathbf{I})$, is a mixture of Gaussians.
\cite{Wu:2010} show that as $M$, the cardinality of $X$, grows, the capacity $C_M$ approaches $\mathcal{C}(\mathrm{snr})=\frac{1}{2}\log(1+\mathrm{snr})$.

\paragraph{Remark 8.} In different practical settings training is sometimes executed with $\Psi_{\mathrm{y}}=\{\mathbf{y}_{i},\mathbf{v}_{i}\}_{i=1}^m$, whereas in other applications the training set consists of the original source-inputs $\Psi_{\mathrm{x}}=\{\mathbf{x}_{i},\mathbf{v}_{i}\}_{i=1}^m$, while the inputs in practice are $\mathbf{y}_{i}$, i.e., perturbed or noisy measurements, and it is expected for the mapping to generalize well. The results in Theorem~\ref{Theorem 4} apply to both cases, although they do not guarantee the performance of the predictor is optimal. This case-study is addressed in Theorem~\ref{Theorem 7} below. 

\begin{theorem}[Generalization error for a noisy input]
\label{Theorem 7}
Let $g(\cdot)$ be a deterministic bijective function. %
Let $\Psi_{\mathrm{x}} = \big\{ \{\mathbf{x}_{i},\mathbf{v}_{i}\}_{i=1}^m : 
\mathbf{x}_i \sim P_X, \quad \mathbf{v}_i= g(\mathbf{x}_i), \ \mathbf{x}_i \in \mathcal{X}^{n\times 1},\ \mathbf{v}_i \in \mathcal{V}^{d\times 1} \big\}$ be a training set that is generated by randomly drawing samples from $P_X$ and labeling them by the target function $g(\cdot)$. Let $h_{\Psi_{\mathrm{x}}} : \mathcal{X}^n \rightarrow \mathcal{V}^d$ be a trained predictor such that, $h_{\Psi_{\mathrm{x}}}(\mathbf{x})=\mathcal{F}(\mathcal{G}(\mathbf{x}))$, where $\mathcal{F}: \mathcal{X}^n \rightarrow \mathcal{Z}^q$, and $\mathcal{G} : \mathcal{Z}^q \rightarrow \mathcal{V}^d$.
Assume the predictor $h_{\Psi_{\mathrm{x}}} : \mathcal{X}^n \rightarrow \mathcal{V}^d$ was trained successfully to yield $\mathcal{L}(h_{\Psi_{\mathrm{x}}}) \leq \Delta << 1$, with $\ell(\cdot)=\|\cdot\|$.  
Let $\Psi_{\mathrm{y}}$ be a test set that is generated by randomly drawing samples from $P_{Y|X}$ and labeling them by the target function such that,  
$\Psi_{\mathrm{y}} = \big\{ \{\mathbf{y}_{i},\mathbf{v}_{i}\}_{i=1}^m : 
\mathbf{y}_i \sim P_{Y|X}(\cdot|\mathbf{x}), \quad \|\mathbf{y}_i-\mathbf{x}_i \| \leq \sigma,\quad \ \mathbf{y}_i \in \mathcal{Y}^{n\times 1},\ \mathbf{v}_i \in \mathcal{V}^{d\times 1}\big\}$. 
Further assume that $\mathcal{F}(\cdot)$ is Lipschitz continuous with constant $K_c$, and that $\mathcal{G}(\cdot)$ is Lipschitz continuous with constant $K_G$.
Then, $\mathcal{L}(h_{\Psi_{\mathrm{x}}}(\mathbf{y})) \leq \Delta + \sigma K_c K_G $.
\end{theorem}
\begin{proof}
By Lipschitz continuity for the embedding we have,
\begin{equation}
\|\mathcal{F}(\mathbf{y}_i)-\mathcal{F}(\mathbf{x}_i)\| \leq  K_c{\|\mathbf{y}_i-\mathbf{x}_i\|} \leq \sigma K_c.
\end{equation}
Therefore,
\begin{equation}
\|\mathcal{G}(\mathcal{F}(\mathbf{y}_i))-\mathcal{G}(\mathcal{F}(\mathbf{x}_i))\| \leq   \sigma K_c K_G.
\end{equation}
And so the generalization error of the predictor $h_{\Psi_{\mathrm{x}}}$ with input $\mathbf{y}$ obeys,
\begin{flalign}
\nonumber
\mathcal{L}(h_{\Psi_{\mathrm{x}}}(\mathbf{y})) & = E_{(y,v) \sim P_{Y,V}}\big[\ell \big(h_{\Psi_{\mathrm{x}}}(\mathbf{y}),\mathbf{v}\big)\big]
=
E_{(x,y) \sim P_{X,Y}}\big[\ell \big(h_{\Psi_{\mathrm{x}}}(\mathbf{y}),\mathbf{v}\big) \big| \mathbf{x}\big]
\\
\nonumber
& 
\leq
E_{y|x \sim P_{Y|X}} E_{x \sim P_{X}}\big[\ell \big(\mathcal{G}(\mathcal{F}(\mathbf{x})),\mathbf{v}\big) + \ell\big(\mathcal{G}(\mathcal{F}(\mathbf{x})),\mathcal{G}(\mathcal{F}(\mathbf{y}))\big)]
\\
\nonumber
& 
\leq
E_{x \sim P_{X}}\big[\ell \big(h_{\Psi_{\mathrm{x}}}(\mathbf{x}),\mathbf{v}\big)] 
+ \sigma K_c K_G \leq \Delta + \sigma K_c K_G, 
\end{flalign}
where we have used that $g(\cdot)$ is a deterministic bijective function, the triangle inequality, and the Lipschitz continuity. 
\end{proof}

\subsection{Representation rate for a compressed output}
\begin{figure}
    \centering
    \includegraphics[width=0.87\textwidth]{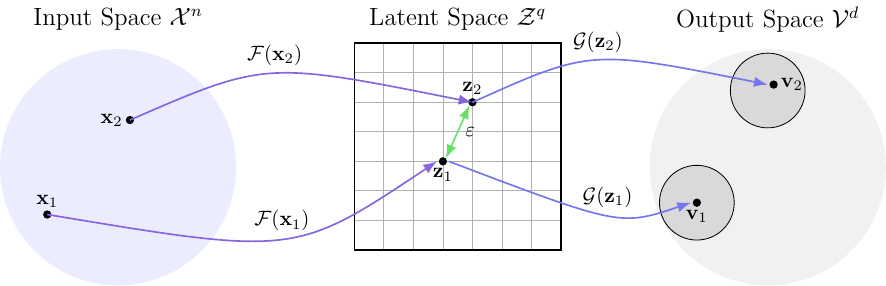}
		\caption{\footnotesize Illustration of a rate-distortion setting.}
		\label{fig3}
\end{figure}

\begin{definition}[Bounded Distortion]
A distortion measure is said to be bounded if the maximum
value of the distortion is ﬁnite:
\begin{equation}
d_{\mathrm{max}} = \max_{\hat{\mathbf{v}}\in \mathcal{V}^d,\mathbf{v}^* \in \mathcal{V}^d} d(\hat{\mathbf{v}},\mathbf{v}^*), 
\end{equation}
Where $\hat{\mathbf{v}}=\mathbf{h}_{\Psi}(\mathbf{x})$ is the model's output, $\mathbf{v}^*$ is the ground truth, and $d(\mathbf{x},\mathbf{y})$ is a distortion measure. 
\end{definition}

\begin{theorem}[Embedding rate-distortion for a bijective mapping]
\label{Theorem 5}
Assume a training set that is generated by randomly drawing samples from $P_X$ and labeling them by the target function $g(\cdot)$,  
$\Psi = \big\{ \{\mathbf{x}_{i},\mathbf{v}_{i}\}_{i=1}^m : 
\mathbf{x}_i \sim P_X, \quad \mathbf{v}_i= g(\mathbf{x}_i), \ \mathbf{x}_i \in \mathcal{X}^{n\times 1},\ \mathbf{v}_i \in \mathcal{V}^{d\times 1} \big\}$. $g(\cdot)$ is a deterministic bijective function. 
For $n$ and $d$ sufficiently large, 
the rate-distortion function of a source $X$, and an output $\hat{V}$ 
with distortion $d(v,\hat{v})$ 
\begin{equation}
\mathcal{R}(D) = \min_{P_{\hat{V}|X}: E [d(v,\hat{v})]\leq D} I(X, \hat{V}) = H(X) -\max_{P_{\hat{V}|X}: E [d(v,\hat{v})]\leq D} H(X|\hat{V})
\end{equation}
is the minimum achievable rate. 
Therefore, 
\begin{equation}
\mathcal{Q}_{\mathrm{z}} \geq n I(X, \hat{V}). 
\end{equation}
\end{theorem}
\textit{Proof.} See Appendix \ref{AppC}.

Note that in a specific setting where $d(\cdot)=\ell(\cdot)$, the expected distortion $E [d(v,\hat{v})]=\mathcal{L}(h_{\Psi})$ is the generalization error. 

In \citep{Pereg:2023A} we defined the terms for the sample complexity to yield a diminishing generalization error.
Here we assumed that the training set is large enough.

We can now combine the two main results that we have proved so far.

\subsection{Representation rate for noisy input and compressed output}

\begin{theorem}[Source-channel representation rate-distortion]
\label{Theorem 6}
Let $(X,V)\sim P_{X,V}$ be an ergodic input-output probability mass function we will refer to as a source. 
A sequence of $n$ input symbols $\mathbf{x} \in \mathcal{X}^{n \times 1}$ is the input of a discrete
channel $P_{Y|X}$. The output of the channel $\mathbf{y} \in \mathcal{Y}^{n \times 1}$  
is mapped onto the reconstruction alphabet $\hat{\mathbf{v}}= h_{\Psi}(\mathbf{y})$, 
$h_{\Psi}:  \mathcal{Y}^{n \times 1} \rightarrow \mathcal{V}^{d \times 1}$.
Let $D = Ed(\mathbf{v},\hat{\mathbf{v}})$ be the average distortion achieved by this combined scheme.
A source with rate-distortion $\mathcal{R}(D)$ can
be sent over a channel of capacity $\mathcal{C}$ and recovered with distortion $D$
if and only if $\mathcal{R}(D) < \mathcal{C}$.
\begin{equation}
\mathcal{R}(D) \leq I(X, \hat{V}) \leq I(X;Y) \leq \mathcal{C}.
\end{equation}
Thus, the representation rate obeys,
\begin{equation}
I(X, \hat{V}) < \frac{\tilde{\mathcal{Q}}_{\mathrm{z}}}{n} < I(X;Y).
\end{equation}
\end{theorem}
\textit{Proof.} See Appendix \ref{AppD}.

\section{Conclusions $\&$ Discussion}

A core question in understanding learning mechanisms relies in understanding whether 
information about the input is fully preserved in the corresponding representation.
In this study, we investigated 
desired properties of signal representation
based on the expected rate of information ``communicated" by the learning model. 
We attempt to formalize trade-offs and to answer the question: What is the capacity of a machine learning model to compute a desired function? \citep{Viterbi:2025}.
As intuitively expected, for a bijective mapping, the representation rate is directly determined by the entropy of the source. 
In a noisy environment, or when there are multiple possible observations of the source input, the maximal non-zero support of the embedding can only be up to the number of distinguishable signals determined by the mutual information between the source and the observed input.
If the representation is compressed, the minimal representation rate for a given distortion, is determined by the mutual information between the source and the compressed output. 
We observed that the constraints on the representation-rate in different settings directly affect the required embedding cardinality and embedding dimensionality. 
Recent work has established that language models are injectives and hence invertible. 
That is, decoder-only transformers almost surely map different
prompts to different hidden states \citep{Nikolaou:2025}. The authors postulate that collisions can never occur under gradient-based training.
In other words, transformer representations are
structurally lossless. 
More often than not, it is assumed that different inputs could collapse to
the same hidden state, making exact recovery of the input impossible. That is, representations discard information that is critical for inversion of the task, which raises concerns for transparency, robustness, and safe deployment, as it suggests that the link between input and its representation is inherently lossy. Several recent works address the non-determinism of LLMs indicating that the primary cause is the non-associative nature of floating-point arithmetic under limited numerical precision \citep{Yuan:2025,He:2025}. 
Here we addressed related settings in a more general framework, revealing inherent limitations of representation learning. 
An interesting direction for future work could empirically investigate the proposed theory and its practical utility for LLMs.

In addition to the theoretical analysis above, our future work will continue to explore and analyze specific examples and applications as well as the practical implications of representation information-theoretic bounds, the resulting  
constraints on architecture's choice, and possibilities for improvements in efficient learning. We noted that the established theorems are independent of the hypothesis class of $h_\Psi(\cdot)$. That is, if the representation rate at the final layer of the encoder is insufficient at the representation level, no architecture can compensate for that. That said, the architecture leading to the final embedding layer and from the embedding to the target may affect the performance of the estimation. Our future work will further investigate these concerns.

\appendix

\section{Proof of the Theorem ~\ref{Theorem 3} - Embedding representation rate - bijective mapping}
\label{AppA}
For simplicity of presentation, we assume throughout the proofs that $n\mathcal{R}$ is an integer.
\begin{proof}
We analyze a representation as shown in Figure~\ref{fig1}.
Recall the model training set is $\Psi = \big\{ \{\mathbf{x}_{i},\mathbf{v}_{i}\}_{i=1}^m : 
\mathbf{x}_i \sim P_X, \quad \mathbf{x}_i= g(\mathbf{v}_i), \ \mathbf{x}_i \in \mathcal{X}^{n\times 1},\ \mathbf{v}_i \in \mathcal{V}^{d\times 1} \big\}$, where $g(\cdot)$ is a deterministic bijective function. 
The proof partially follows the outline of Shannon coding theorem \citep{Kramer:2008}. 
We assume paired ground truth signals $\{\mathbf{x}_i, \mathbf{v}_i\}_{i=1}^{M}$, $M=2^{n\mathcal{R}}$. 
Denote the original ``message" $w$, drawn from the index set $\{1, 2,...,M\}$, which results in the input
signal $\mathbf{x}_w$ that is received by the model (neural network) as a random sequence. 
That is, the input to the learning model is $\mathbf{x}_w$. 
The neural net's encoder then yields the representation $\mathbf{z}=\mathcal{F}(\mathbf{x})$. 
The decoder obtains the output $\hat{\mathbf{v}}=\mathcal{G}(\mathbf{z})$ that corresponds with the pair $\{\mathbf{x}_i,\mathbf{v}_i\}$ with index $i=\hat{w}$. The model makes an error if $\hat{w}$ is not the same as the index $w$ that was ``transmitted", that is if $\hat{\mathbf{v}}\neq\mathbf{v}$. 

Let $\mathbf{z}$ be an embedding space with $\mathcal{Q}_{\mathrm{z}} \geq n (H(X) + \epsilon_\mathrm{z})$.
In this case $|\mathcal{Z}|^q=2^{\mathcal{Q}_{\mathrm{z}}} \geq 2^{n (H(X) + \epsilon_\mathrm{z})}$.
Thus, by Theorem~\ref{Theorem 1.2}, for $n$ large enough and $\epsilon_\mathrm{z} \geq \epsilon$, we have $|\mathcal{Z}|^q \geq |A^n_\epsilon(P_X)|$.
Therefore, provided that $d \log_2|\mathcal{V}| \geq 2^{nH(\mathbf{x})}$, that is
by assumption, for every $\mathbf{x} \in A^n_\epsilon(P_X)$ there is a corresponding unique output $\mathbf{v}$. 
Hence $\forall \mathbf{x}\in A^n_\epsilon(P_X) \ \exists \ \mathbf{z} \in \mathcal{Z}^q$ such that $h_\Psi(\mathbf{x})=\mathbf{v}^*$, and $P_e=Pr[ x^n \notin A^n_\epsilon(P_X)]\rightarrow 0$, where $\mathbf{v}^*$ denotes the ground truth value at the predicted output.
Denote $A=\{x^n : x^n \in  A^n_\epsilon(P_X) \}$, $B=\{x^n : x^n \notin  A^n_\epsilon(P_X) \}$, such that the generalization error (risk) over $x^n \in A$ is $\varepsilon^\mathcal{F}_A=0$. 
However if $x^n \notin A^n_\epsilon(P_X)$ we can assume some unknown output $\hat{v}^d$ (the encoder sends to the decoder some unknown $z^q$ generated by the trained learning system), with error $\varepsilon^\mathcal{F}_B$.
\begin{equation}\label{A4.2}
\mathcal{L}(h_{\Psi})   \leq Pr[ x^n \in A^n_\epsilon(P_X) ] \varepsilon^\mathcal{F}_A + Pr[ x^n \notin A^n_\epsilon(P_X) ] \varepsilon^\mathcal{F}_B.
\end{equation}
\begin{equation}\label{A4.3}
\mathcal{L}(h_{\Psi})  \leq Pr[ x^n \in A^n_\epsilon(P_X) ] \varepsilon^\mathcal{F}_A \rightarrow 0.
\end{equation}

Let $\mathbf{z}$ be an embedding space with $\mathcal{Q}_{\mathrm{z}} < n (H(X) - \epsilon_\mathrm{z})$.
We have
\begin{flalign*}
\nonumber
1-P_e & = \sum_{ \mathbf{x} \ : \ h_\Psi(\mathbf{x})=\mathbf{v}^* } P_X(\mathbf{x}) 
\\
\nonumber
& =
\sum_{\mathbf{x} \in A^n_\epsilon(P_X) \cap h_\Psi(\mathbf{x})\in P_{V|X}} P_X(\mathbf{x}) +
\sum_{\mathbf{x} \notin A^n_\epsilon(P_X) \cap h_\Psi(\mathbf{x})\in P_{V|X}} P_X(\mathbf{x}) 
\\
\nonumber
&
\leq
|\mathcal{Z}|^q \max_{\mathbf{x} \in A^n_\epsilon(P_X)}{P_X(\mathbf{x})}+
Pr[\mathbf{x} \notin A^n_\epsilon(P_X)]  
\\
\nonumber
&
\leq
2^{n(H(x)-\epsilon_\mathrm{z})} 2^{-n(H(X)-\epsilon)} + Pr[\mathbf{x} \notin A^n_\epsilon(P_X)]  
\end{flalign*}
For n large enough the right term approaches zero, and for $\epsilon_\mathrm{z}-\epsilon>0$ the first term approaches zero. Therefore $P_{e}\rightarrow 1$. 
\end{proof}

\section{Proof of Theorem ~\ref{Theorem 4} - Representation Capacity}
\label{AppB}
\begin{proof}

We analyze a representation as shown in Figure~\ref{fig2}.
The proof follows the outline of achievability of Shannon channel coding theorem \citep{Kramer:2008}. 
We assume paired ground truth signals $\{\mathbf{x}_i, \mathbf{v}_i\}_{i=1}^{M}$, $M=2^{n\mathcal{R}}$. 
Denote the original ``message" $w$, drawn from the index set $\{1, 2,...,M\}$, which results in the
signal $\mathbf{x}_w$ that goes through a system (aka channel) (noise, degradation, quantization...) and is received by the model (neural network) as a random sequence $\mathbf{y} \sim P_{Y|X}(\cdot|\mathbf{x})$. The input to the learning model is $\mathbf{y}$. 
The neural net's encoder then yields the representation $\mathbf{z}=\mathcal{F}(\mathbf{y})$. 
The decoder obtains the output $\hat{\mathbf{v}}$ that corresponds with the pair $\{\mathbf{x}_i,\mathbf{v}_i\}$ with index $i=\hat{w}$. 
That is, $\hat{\mathbf{v}}=h_{\Psi}(\mathbf{y})=\mathcal{G}(\mathbf{z})$. 
The model makes an error if $\hat{w}$ is not the same as the index $w$ that was ``transmitted", that is if $\hat{\mathbf{v}}\neq\mathbf{v}$. 
Cost constraint: we further impose a cost constraint over the average cost function $E s(\mathbf{x},\mathbf{y}) \leq S_{\mathrm{max}}$. 

\begin{definition} An $(M, n)$ code for the channel $(X,P_{Y|X}(\mathbf{y}|\mathbf{x}), Y)$ and the model $h_{\Psi}=(\mathcal{F}(\mathbf{y}), \mathcal{G}(\mathbf{z}))$ consists of
the following:
\begin{enumerate}
\item An index set $\{1, 2,...,M\}$.
\item An encoding function $X^n : \{1, 2,...,M\} \rightarrow \mathcal{X}^n$, yielding codewords
$x^n(1), x^n(2), ..., x^n(M)$. The set of codewords is called the codebook.
\item
A decoding representation
$\mathcal{F} : \mathcal{Y}^n \rightarrow \mathcal{Z}^q$
which is a deterministic rule that assigns a representation to each possible
received vector.
\item
A decoding map $\mathcal{G} : \mathcal{Z}^q \rightarrow \mathcal{V}^d$
which is a deterministic rule that assigns an output to each representation.
\end{enumerate}
$h_{\Psi}$ is the concatenation of $\mathcal{F}$, and $\mathcal{G}$. 
\end{definition}

Note that in our setting the learning model that maps $y^n$ to $z^q$ and then maps $z^q$ to $v^d$ (encoder-decoder in machine learning literature) is corresponding with the decoder in information theory literature discussing channel capacity. The encoding function corresponds with any heuristic stage determining the original ground truth which corresponds with the noisy or distorted measurement that is the input to the neural net from a set of indexes of possible inputs.
This stage can also correspond with any preprocessing stage applied to the signal, which yields the input to the neural net, such as: rescaling, normalization, clipping and quantization.  
The goal is to find the maximum rate $\mathcal{R}$ for which $Pr[\hat{w} \neq w]$ is arbitrarily close to zero and thus find the required $\mathcal{Q}_{\mathrm{z}}$ for the representation.

Similarly to \citep{Pereg:2023A}, in this case, we assume $P_{Y|X}(\mathbf{y}|\mathbf{x})$, e.g., $y=u(x)+e$, such that each possible $y$ input induces a probability mass function over the possible $x$'s. By assumption, the model efficiently learns the mapping $h_{\Psi}: \mathbf{y} \rightarrow \mathbf{v}$, and therefore the corresponding ground truth $x$'s are distinguishable.
In other words, the same $y$ is not associated with two different $x$'s. Since we assumed a surjective mapping $f: \mathcal{Y}^n \rightarrow \mathcal{X}^n$, for a distinguishable subset of input sequences, there exists only one $x^n$ that could have caused a particular $y^n$ with high probability. We can therefore reconstruct the
sequence at the output with a negligible probability of error, by
mapping the observations into the proper ``widely spaced'' hidden sequences.
We can define the conditional entropy $H(Y|X)$ assuming they are ergodic and have a stationary coupling \citep{Gray:2009}.
Defining their mutual information $I(X;Y)=H(Y)-H(Y|X)$, their jointly typical
set follows similar properties \citep{ThomasCover:2006}.
Define $B=\{(x^n,y^n): (x^n,y^n) \in A^n_\epsilon(P_{X,Y}), y^n \in A^n_\epsilon(P_{Y}) , x^n \in A^n_\epsilon(P_{X}) \}$,and $A^n_\epsilon(P_{X,Y}|x^n)= \{y^n : (x^n,y^n) \in A^n_\epsilon(P_{X,Y})\}$.
Observe that $A^n_\epsilon(P_{X,Y}|x^n)=\emptyset$ if $x^n \notin A^n_\epsilon(P_{X})$  \citep{Kramer:2008}.
For $n$ sufficiently large and small $\epsilon$,
\begin{equation}\label{C3.1}
\mathrm{Pr}[Y^n \in A^n_\epsilon(P_{X,Y}|x^n)|X^n=x^n] = 1.
\end{equation}
\begin{equation}\label{C3.2}
\mathrm{Pr}[(x^n,y^n) \in B] = 2^{-nI(X;Y)}.
\end{equation}
\begin{equation}\label{C3.3}
|B| = 2^{nI(X;Y)}.
\end{equation}
Roughly speaking, if $y^n$ and $x^n$ are jointly typical, then we can resolve an input $y^n$ as $x^n$ and infer the correct $v^d$. There are approximately $2^{nH(Y|X)}$ equally probable $y^n$ sequences, for each typical output sequence $x^n$. We assume
that no two $x^n$ sequences correspond to the same $y^n$ output sequence,
otherwise, the learner will not be able to decide which $x^n$ sequence it originated from.
There are approximately $2^{nH(Y)}$ possible typical $y^n$ sequences. This set
is split into sets of size $2^{nH(Y|X)}$,
associated with different $x^n$ sequences. Therefore, the total number of distinguishable sets is less than or equal
to {\small $2^{n(H(Y)-H(Y|X))}=2^{nI(X;Y)}$}.
Hence, we can have at most $2^{nI (X;Y)}$ disjoint sequences of length $n$.

An achievable rate: 
Code construction: generate $2^{n\mathcal{R}}$ code words $x^n(w)$, 
$w=1,2,...,n$ using $P_{X^n}(x^n)$.
Encoder: Given $w$, $x^n(w)$ is ``transmitted".
Decoder: Given $y^n$ output $z^q(\tilde{w})$ and consequently $v^d(\tilde{w})$ such that $(x^n(\tilde{w}),y^n)\in A_{\epsilon}^n(P_{X,Y})$. 
If there is one or more such $\tilde{w}$, then choose one as $\hat{w}$.
If there is no such $\hat{w}$, then output $z^q(\hat{w}=1)$. 
\begin{enumerate}
\item Suppose that $x^n(w) \notin A^n_{\epsilon}(P_{X})$, 
when n is large $Pr[x^n(w) \notin A^n_{\epsilon}(P_{X})]=0$.

\item Suppose that $x^n(w) \in A^n_{\epsilon}(P_{X})$ but $(x^n(w),y^n) \notin A^n_{\epsilon}(P_{X,Y})$, 
when $n$ is large $\mathrm{Pr}[Y^n \in A^n_\epsilon(P_{X,Y}|x^n)|X^n=x^n]=0$. %

\item Suppose $(x^n(w),y^n) \in A^n_{\epsilon}(P_{X,Y})$, but we also find a $\tilde{w} \neq w$ such that 
$(x^n(\tilde{w}),y^n) \in A^n_{\epsilon}(P_{X,Y})$

\begin{flalign}
\nonumber
P_e(w) & = Pr \Bigg[ \bigcup_{\tilde{w} \neq w }  \big\{ (x^n(w),y^n) \in A^n_{\epsilon}(P_{X,Y})|x^n(w) \in A^n_{\epsilon}(P_{X}) \big\}  \Bigg] 
\\
\nonumber
&
\leq \sum_{\tilde{w} \neq w } Pr\big[ (x^n(w),y^n) \in A^n_{\epsilon}(P_{X,Y})|x^n(w) \in A^n_{\epsilon}(P_{X})   \big] 
\\
& \leq \big( 2^{n\mathcal{R}}-1 \big) 2^{-n[I(X;Y)-2 \epsilon H(Y)]}.
\end{flalign}
The first inequality follow from the union bound, the second inequity follows from Theorem 1.3 in \cite{Kramer:2008} .   
Implying that for a large $n$ we can choose
\begin{equation}
\mathcal{R} < I(X;Y)-2 \epsilon H(Y)
\end{equation}
to drive $P_e(w)$ to zero.
\end{enumerate}

We compute the average cost for $(x^n,y^n) \in A^n_{\epsilon}(P_{X,Y})$:
\begin{equation}
E s(\mathbf{x},\mathbf{y})= \sum_{(x^n,y^n) \in B} \mathrm{Pr}[(x^n,y^n) \in B] s(x^n,y^n) 
\leq S_{\mathrm{max}}
\end{equation}

Combining the above results, there is a predictor $h_{\Psi}$ that approaches the rate
\begin{equation}
\mathcal{C}(S) = \max_{P_x(\mathbf{x}):E s(\mathbf{x},\mathbf{y})\leq S_{\mathrm{max}}} I(X;Y)
\end{equation}
$\mathcal{R}<I(X;Y)$ imposes a constraint on the required effective support of the embedding: $\mathcal{Q}_{\mathrm{z}} < nI(X;Y)$. If the support size is larger, i.e., $\mathcal{Q}_{\mathrm{z}} \geq nI(X;Y)$, but the non-zero support still obeys $\tilde{\mathcal{Q}}_{\mathrm{z}} < nI(X;Y)$ then the predictor's error approaches zero.

\paragraph{Proof of converse.} 

If $\mathcal{R} > I(X;Y)$, in the sense that we are attempting to resolve the observation to more than the possible distinguishable $\mathbf{x}$'s, there exists $\epsilon>0$ such that $P_e(w)>0$ for every predictor $h_{\Psi}$.
Alternatively, for every predictor with $P_e \triangleq Pr[h_{\Psi} (\mathbf{y}) \neq \mathbf{v}] \rightarrow 0, \ \forall \mathbf{y} \sim P_{Y|X}$, the rate must satisfy $\mathcal{R} \leq I(X;Y)$. 
The proof of the converse uses Fano’s inequality and basic properties of mutual information.
\begin{flalign}
\nonumber
n\mathcal{R} & = H(W) = H(W|Y^n)+I(W;Y^n) 
\\
\nonumber
&
\leq H(W|Y^n)+I(X^n;Y^n)  \leq H(W|\hat{W}) + I(X^n;Y^n) 
\\
& \leq P_e n \mathcal{R} + 1 + n I(X;Y),
\end{flalign}
where we assume $|\mathcal{W}|$ is at most $2^{n\mathcal{R}}$, $W-X^n-Y^n-\hat{W}$ is a Markov chain, and that the limit $I(X;Y)=\lim_{n\rightarrow\infty}\frac{1}{n}I(X^n,Y^n)$ exists for stationary ergodic processes \citep{ThomasCover:2006}. 
To derive the last in equality, we used Fano's inequality \citep{Kramer:2008}, namely $H_2(P_e)+P_e\log_2(|\mathcal{W}|-1) \geq H(W|\hat{W})$. 
Therefore, we have for large $n$, and a predictor $h_{\Psi}$ that yields $P_e \rightarrow 0$, then the rate is $\mathcal{R} \leq I(X;Y)$.
\end{proof}

\section{Proof of the Theorem ~\ref{Theorem 5} - Representation rate-distortion}
\label{AppC}

\begin{proof}

We analyze a representation as shown in Figure~\ref{fig3}.
The proof follows the outline of achievability of rate-distortion region \citep{Kramer:2008}. 
Rate distortion theory is concerned with quantization or lossy compression. 
Here we consider a different setting. 
A source $P_X(\cdot)$ with alphabet $\mathcal{X}$ emits a sequence $x^n$ that is passed to an
encoder. The encoder $\mathcal{F} : \mathcal{X}^n \rightarrow \mathcal{Z}^q$ maps the input sequence to an embedding space of 
$|\mathcal{Z}|^q=2^{n\mathcal{R}}$ possible sequences $z^q$. 
The decoder $\mathcal{G} : \mathcal{Z}^q \rightarrow \hat{\mathcal{V}}^d$ maps $z^q$ into $\hat{v}^d$.
The goal is to ensure that the average non-negative
and real-valued distortion is bounded $E [d(v^d,\hat{v}^d)] \leq D$.
We further assume that the maximum distortion is finite, i.e., $d(v^d,\hat{v}^d) \leq d_{max}$.

We define here the joint typical set, for $n$ and $d$ large enough,
\begin{flalign}
\nonumber
A^{n,d}_{\epsilon}(P_{X,V}) =
& \bigg\{(x^n,v^d) \in \mathcal{X}^n \times \mathcal{V}^d : 
\Big| -\frac{1}{n} \log_2 P_{X,V}(x^n,v^d) - H(X,V) \Big| < \epsilon 
\\
\nonumber
&
x^n \in \mathcal{X}^n : 
\Big| -\frac{1}{n} \log_2 P_X(x^n) - H(X) \Big| < \epsilon 
\\
&
v^d \in  \mathcal{V}^d : 
\Big| -\frac{1}{d} \log_2 P_V(v^d) - H(V) \Big| < \epsilon 
\bigg\}
\end{flalign}

An achievable rate. 
Given $x^n$, output $\hat{v}^d$ such that $(x^n,\hat{v}^d) \in A^{n,d}_{\epsilon}(P_{X,V})$.
To bound $E d(v^d,\hat{v}^d) $ we partition the sample space into three events:
\begin{equation}
\mathcal{E}_1 = \{ x^n \notin A^n_{\epsilon}(P_{X}) \},
\end{equation}
\begin{equation}
\mathcal{E}_2 = \mathcal{E}^c_1 \bigcap 
\Bigg\{ \bigcap^{2^{n\mathcal{R}}}_{i=1} \big\{ (x^n,\hat{v}^d_i) \notin A^{n,d}_{\epsilon}(P_{X,\hat{V}}) \big\} \Bigg\},
\end{equation}
\begin{equation}
\mathcal{E}_3 = (\mathcal{E}_1 \cup \mathcal{E}_2)^c,
\end{equation}
where $\mathcal{E}^c_1$ denotes the complement of $\mathcal{E}_1$.  
Next, we apply the Theorem of Total Expectation,
\begin{equation}
E[d(v^d,\hat{v}^d)] = \sum^3_{k=1} Pr[\mathcal{E}_k] E[ d(v^d,\hat{v}^d) | \mathcal{E}_k].
\end{equation}

\begin{enumerate}
\item Suppose that $x^n(w) \notin A^n_{\epsilon}(P_{X})$, 
the upper bound on the average distortion is $d_{\mathrm{max}}$, but when $n$ is large $Pr[x^n(w) \notin A^n_{\epsilon}(P_{X})]=0$. 
\item Suppose that $x^n\in A^n_{\epsilon}(P_{X})$ but non of the possible outputs given the representation cardinality is in the joint typical set, that is, $(x^n,\hat{v}_i^d) \notin A^{n,d}_{\epsilon}(P_{X,\hat{V}}) \ \forall
i=1,2,...2^{n\mathcal{R}}$. Thus,
\begin{flalign}
\nonumber
P_e & =  Pr \Bigg[ \bigcup_{i=1}^{2^{n\mathcal{R}}}  \big\{ (x^n,\hat{v}^d_i) \notin A^{n,d}_{\epsilon}(P_{X,\hat{V}}) | x^n\in A^n_{\epsilon}(P_{X}) \big\}  \Bigg] 
\\
\nonumber
&
=
[1-Pr[(x^n,\hat{v}^d_i) \in A^{n,d}_{\epsilon}(P_{X,\hat{V}}) | x^n\in A^n_{\epsilon}(P_{X}) ]]^{2^{n\mathcal{R}}}
\\
\nonumber
&
\approx [1-2^{-nI(X;\hat{V})}]^{2^{n\mathcal{R}}}
\\
&
\leq
\exp(-2^{n(\mathcal{R}-I(X;\hat{V}))}), 
\end{flalign}
since $(1-x)^m \leq e^{-mx}$.
For a large n we can choose
\begin{equation}
\mathcal{R} > I(X;\hat{V})
\end{equation}
to drive $P_e$ to zero.
The upper bound on the average distortion is again $d_{\mathrm{max}}$.
\item Suppose $(x^n,\hat{v}^d) \in A^{n,d}_{\epsilon}(P_{X,\hat{V}})$, the distortion is, by assumption
\begin{flalign}
E d(v^d,\hat{v}^d)  = \sum_{x^n \in A^n_{\epsilon}(P_{X})} P(x^n) d(v^d,h_{\Psi}(x^n)) \leq D.
\end{flalign}
\end{enumerate}
Overall,
\begin{equation}
E d(v^d,\hat{v}^d)  \leq D + P_e d_{\mathrm{max}}.
\end{equation}
And for $\mathcal{R} > I(X;\hat{V})$
\begin{equation}
E d(v^d,\hat{v}^d)  \leq D.
\end{equation}

\paragraph{Proof of converse.} 
We need to show that for any source $X^n$ with $2^{n\mathcal{R}}$ sequences with $\lim_{n\rightarrow \infty} \sup E d(v,\hat{v}) \leq D$,
we must have $\mathcal{R} \geq \mathcal{R}(D)$. 
The representation rate is $\mathcal{R}$ bits per input symbol.
Therefore we have $n\mathcal{R}$ bits per model input which could represent $2^{n\mathcal{R}}$ sequences (messages, words, patches, etc.) at most. Therefore \footnote{We denote $H(X^n) \triangleq E[-\log P_X(x^n)]$, where $H(X)=\lim_{n \rightarrow \infty} \frac{1}{n} H(X^n)$ \citep{ThomasCover:2006}}, $ H(\hat{V}^d) \leq \log 2^{n\mathcal{R}}$
.
Thus,
\begin{equation}
n\mathcal{R} \geq  H(\hat{V}^d) \geq  H(\hat{V}^d) -  H(\hat{V}^d|X^n) = I(X^n;\hat{V}^d) =  H(X^n) -  H(X^n|\hat{V}^d).
\end{equation}
Assuming that the limit $I(X;V)=\lim_{n\rightarrow\infty}\frac{1}{n}I(X^n,V^d)$ exists for stationary ergodic processes \citep{ThomasCover:2006},
\begin{equation}
\mathcal{R} \geq \min_{P_{\hat{V}|X}: E [d(v,\hat{v})] \leq D} I(X, \hat{V}) = \mathcal{R}(D).
\end{equation} \end{proof}

\section{Proof of the Theorem ~\ref{Theorem 6} - Representation rate source-channel separation}
\label{AppD}
\begin{proof}
The proof follows similar outline to the proof of Source–Channel Separation Theorem \\ \citep{Gamal:2011}.
We use a heuristic of separate lossy source coding and channel coding, although in practice in our setting there is no separation in the prediction stage. %

\paragraph{Proof of achievability.} 
In Theorem~\ref{Theorem 5} we have established that for any $\epsilon> 0$ and for an input-output sequence $\{\mathbf{x}_i, \mathbf{v}_i\}\sim P_{X,V}$ there exists an estimator $h_\Psi$ that yields a lossy representations with rate
$\mathcal{R}(D)$ with an expected distortion less than or equal to $D$. 
We treat the index $i$ for each sequence $\mathbf{x}_i$ as a message to be sent over the channel $P_{Y|X}$.
By Theorem~\ref{Theorem 4} the source sequence can be reliably mapped to $\mathbf{v}_i$ from an input $\mathbf{y}_i\sim P_{Y|X}(\cdot|\mathbf{x})$ if $\mathcal{R}(D) \leq \mathcal{C}$.
The estimator predicts the reconstruction sequence corresponding with the received
index. If the estimator makes an error, the distortion is upper bounded by $d_{\mathrm{max}}$.
Because the probability of error tends to zero as $n \rightarrow \infty$, the overall expected distortion
is less than or equal to $D$.

\paragraph{Proof of the converse.}
We wish to show that if a sequence of codes achieves the rate–
distortion, then $\mathcal{R}(D) \leq \mathcal{C}$. 
By the converse proof of the lossy representation-distortion theorem, we know that
\begin{equation}
\mathcal{R}(D) \leq I(X, \hat{V}).
\end{equation}
Now, by the data processing inequality,
\begin{equation}
I(X, \hat{V}) \leq I(X;Y).
\end{equation}
Following similar steps to the converse proof for embedding capacity, we therefore have
\begin{equation}
I(X, \hat{V}) \leq I(X;Y) \leq \mathcal{C}.
\end{equation}
Combining the above inequalities completes the proof of the converse.
\end{proof}

\bibliography{SSL_Pereg_2025}

\end{document}